# Phase evolution and thermal stability of high Curie temperature BiScO$_3$-PbTiO$_3$-Pb(Cd$_{1/3}$Nb$_{2/3}$)O$_3$ ceramics near MPB

Zhencheng Lan[1], Kaiyuan Chen[1], Xiafeng He[1], Shuo Zhou[1], Xiande Zheng[1], Jia Liu[2], Liang Fang[1], Xiuyun Lei[1], Dawei Wang[2*], Biaolin Peng[3*], Laijun Liu[1*]

[1]College of Materials Science and Engineering, Guilin University of Technology, Guilin, China

[2]School of Microelectronics and State Key Laboratory for Mechanical Behaviour of Materials, Xi'an Jiaotong University, Xi'an, China

[3]School of Physical Science & Technology and Guangxi Key Laboratory for Relativistic Astrophysics, Guangxi University, Nanning, China

**Abstract:** Piezoelectric and ferroelectric ceramics with a high Curie temperature ($T_c$) have attracted a growing attention owning to their applications under severe environments. In this work, phase structure, dielectric, ferroelectric and piezoelectric properties of (0.975-$x$)BiScO$_3$-$x$PbTiO$_3$-0.025Pb(Cd$_{1/3}$Nb$_{2/3}$)O$_3$ ceramics ($x$ = 0.58-0.64) were studied. A composition-induced structural transformation occurs from rhombohedral phase to tetragonal phase through an intermediate monoclinic phase with the increasing PT concentration. The relationship between structure and electrical properties of the system were discussed. The BS-$x$PT-PCN system near the morphotropic phase boundary ($x$ = 0.62) exhibits excellent piezoelectric and ferroelectric performances with $d_{33}$ = 508 pC/N, $k_p$ = 56%, and $P_r$ = 40 μC/cm$^2$. The high-temperature piezoelectricity of the sample with MPB ($x$ = 0.62) was characterized by an *in situ* XRD. The excellent thermal stability of the crystal structure and the piezoelectric property indicate that the BS-$x$PT-PCN system is a promising candidate for high temperature piezoelectric applications.

*Corresponding authors.
E-mail addresses: dawei.wang@xjtu.edu.cn (Prof. D. Wang), ljliu2@163.com (Prof. L. Liu).

# 1. Introduction

Materials with excellent piezoelectricity and decent Curie temperature ($T_c$) are broadly needed for varied applications, for example underwater sonars, automotive, aerospace crafts, etc., [1,2] where they are supposed to function in harsh environments. Nevertheless, majority of the piezoelectrics with excellent piezoelectric properties have a fairly low $T_c$. For example, $Pb(Zn_{1/3}Nb_{2/3})O_3$-$PbTiO_3$ and $Pb(Mg_{1/3}Nb_{2/3})O_3$-$PbTiO_3$ single crystals have an extremely large piezoelectric constant ($d_{33}$) of approximately 2500 pC/N [3,4,5,6,7] while a low $T_c$ of about 140-170 °C and an even lower $T_d$ of around 60-120 °C, which limits its application temperature range. Consequently, it is an urgent task to find materials with large $T_c$ and excellent performances to adapt to extreme environments.

In the hunt for exquisite perovskite oxides with excellent performances, solid solutions with a morphotropic phase boundary (MPB) have been considered as the most promising candidates. $PbZr_{1-x}Ti_xO_3$ (PZT) ceramics presents the largest dielectric and piezoelectric properties around the MPB for $x=0.48$. Noheda *et al.* found an medium monoclinic (*Cm*) phase near MPB of PZT [8] as a structural bridge between the rhombohedral (*R3m*) phase and tetragonal (*P4mm*) phase. This finding created a range of structural investigations on the nature of the morphotropic phase boundary [9,10,11,12]. Involving studies of the local structure which have illustrated the theory of the decent piezoelectric performances [13]. In spite of the connection between the crystal structure and piezoelectric performance around the MPB has not been adequately comprehend, the low-symmetry phase is acknowledged to play a vital role in the presence of large piezoelectric performances both in terms of extrinsic [14] and intrinsic contributions [15].

Recently, the $(1-x)$BiScO$_3$-$x$PbTiO$_3$ (BS-PT) solid solution have been thoroughly studied, which showed large piezoelectric coefficient ($d_{33}$ = 460 pC/N) and high Curie temperature ($T_c$ > 450 °C) for the compositions around the MPB ($x$ = 0.64) [16]. Nevertheless, scandia is too costly to limit the commercial manufacture. Based on Bi-based perovskite piezoelectric ceramics, a number of investigator have take care of Bi(Sc$_{3/4}$In$_{1/4}$)O$_3$-PbTiO$_3$, Bi(Mg$_{1/2}$Ti$_{1/2}$)O$_3$-PbTiO$_3$, Bi(Sc$_{3/4}$Co$_{1/4}$)O$_3$-PbTiO$_3$, Bi(Sc$_{1-y}$Fe$_y$)O$_3$-PbTiO$_3$, and Bi(Sc$_{1-y}$Ga$_y$)O$_3$-PbTiO$_3$ systems and achieve some satisfactory piezoelectric properties with the $T_c$ beyond 430 °C [17, 18, 19, 20, 21]. Meanwhile, several researchers attempt to import a third component into the BS-PT binary system to decrease scandia content, for instance, LiNbO$_3$, LiTaO$_3$ [22, 23]. The desirable piezoelectric performances were obtained but the $T_c$ reduced significantly, confining their applications at high temperatures. It is known that Pb(Cd$_{1/3}$Nb$_{2/3}$)O$_3$ is a classic relaxor ferroelectric material. Previous work [24] indicated that Pb(Cd$_{1/3}$Nb$_{2/3}$)O$_3$ has two temperatures of dielectric anomaly: the low-temperature one appears at ~33 °C while the high-temperature anomaly happens at about 387 °C. The composition 0.86PbTiO$_3$-0.14Pb(Cd$_{1/3}$Nb$_{2/3}$)O$_3$, with some appropriately additives, such as MnO$_2$, WO$_3$, and NiO, shows decent electrical performances and the $T_c$ is exceed 450 °C [25]. Herein, it is possible to importe Pb(Cd$_{1/3}$Nb$_{2/3}$)O$_3$ into the BS-PT system to improve the piezoelectric performances and remain high $T_c$.

In this work, the phase structure, dielectric, ferroelectric and piezoelectric properties of the PCN-modified BS-PT systems are investigated by modifying the ratio between BS and PT. Since phase stability is vital consideration for high-temperature piezoelectric ceramics, the structure evaluation as a function of temperature was involved to illustrate the nature of the high-temperature piezoelectricity.

## 2. Experimental procedures

The (0.975-$x$)BiScO$_3$-$x$PbTiO$_3$-0.025Pb(Cd$_{1/3}$Nb$_{2/3}$)O$_3$ ceramics (BS-$x$PT-PCN, $x$ = 0.58, 0.60, 0.62, 0.64) were synthesized via the conventional ceramic technology. Analytical grades Sc$_2$O$_3$ (99.9%), Bi$_2$O$_3$ (99.9%), TiO$_2$ (99.99%), PbO (99.9%), CdO (99.99%), as well as Nb$_2$O$_5$ (99.99%) were selected as the original oxides and were weighed according to the stoichiometric ratio with an extra 2% Bi$_2$O$_3$ and PbO to compensate the volatilization of Bi as well as Pb during the sintering stage. The starting materials was ball milled using ethyl alcohol for 12 h, calcined at 800 °C for 4 h in a covered alumina crucible and milled for an added 12 h to make powders uniform and fine. The dried powder was pressed into 10 mm disks at a pressure of 350 MPa. PVA (polyvinyl alcohol) was combusted off at 650 °C for 4 h, then the pellets were surrounded in calcined powder with the uniform composition and sintered at 1080~1100 °C for 2 h in covered crucible.

The phase structure of BS-$x$PT-PCN was determined via the X-ray diffractometer (XRD, PANalytical, X'Pert PRO) using Cu $K_\alpha$ radiation from RT to 450 °C. Raman spectroscopy was identified with a 532 nm laser and a valid power of 5.0 mW. The micromorphology was measured via scanning electron microscopy (SEM, JSM6380-LV, Tokyo, Japan). After the specimens were poled for 30 min in silicon oil bath under a DC field of 70 kV/cm at 120 °C, $d_{33}$ was tested by a piezoelectric $d_{33}$ meter (ZJ-3D, Institute of Acoustics, Beijing, China). Ferroelectric hysteresis loops were collected via Raniant Precision 10 kV HVI-SC analyzer. The dielectric properties dependence on temperature was detected through an impedance analyzer (Agilent 4294A, USA).

## 3. Results and discussion

3.1 Structure and phase analysis

Figure 1(a) presents XRD patterns of BS-$x$PT-PCN with different PT contents. All the samples were well crystallized as well as revealed a single perovskite structure no distinctly secondary phases, illustrating that PCN was completely solved into BS-PT lattice. Meanwhile, it is found that the crystalline structure of BS-$x$PT-PCN ceramics transforms from rhombohedral to tetragonal as the PbTiO$_3$ content increases. Typical rhombohedral phase can be observed for $x$ = 0.58. As $x$ achieves 0.64, the reflection peak (200) has completely split into (002) and (200) peaks, and then the ceramic is a tetragonal phase, as presented in Figure 1(b). The MPB can be identified at $x$ = 0.62 parting the rhombohedral phase and tetragonal phase. To further approve the result, the $k_p$ and $d_{33}$ depend on PbTiO$_3$ content are presented in Figure 2. It can be clearly seen that the $d_{33}$ and $k_p$ intimately depend on PbTiO$_3$ content, achieving the peak value at $x$ = 0.62, which are 56% and 508 pC/N, respectively. The optimal piezoelectric performances are acquired at the composition close to the morphotropic phase boundary (MPB), which presences a sudden structural transformation [26]. Combing with the XRD results, it is considered that the morphotropic phase boundary (MPB) content locates in the vicinity of $x = 0.62$. As reported, the composition of MPB in the pure $BiSO_3$-$PbTiO_3$ ceramics was around the $PbTiO_3$ content of 0.64 [16]. Obviously, doping with PCN in BS-PT ceramics lead to the MPB composition shifting to lower $PbTiO_3$ content.

To obtain the structure of the BS-$x$PT-PCN systems with $x = 0.58$, 0.60, 0.62 and 0.64, the Rietveld refinements is performed as shown in Figure 3. Details of the refined results were cataloged in Table 1. The samples with $x = 0.58$ and $x = 0.60$ are under the threshold boundary of phase coexistence (resulting in the MPB) and the symmetry is rhombohedral with the space group $R3m$. For the composition at $x = 0.64$, a single tetragonal phase with space group $P4mm$ was used to refine the XRD pattern, resulting in good agreement.

On the other hand, phase coexistence was clearly observed for the composition $x = 0.62$. The best $R$ factors were obtained in the case of a heterogeneous model embracing tetragonal $P4mm$ and monoclinic $Cm$ phase (Table 1, Figure 3). Such presence of an intermediate monoclinic phase, which divides the rhombohedral and the tetragonal region on the $x$-T phase diagram, was first proposed for the by Noheda *et al* [8] for PZT ceramics. Later, a similar phase was discovered during the investigation of the BS-PT [27, 28] ceramics. Given these pioneering work, the presence of the monoclinic phase and the existence of MPB in the ternary system BS-$x$PT-PCN is expected.

Figure 4(a) presents Raman spectra taken at room temperature for the BS-$x$PT-PCN samples with $x$ = 0.58, 0.60, 0.62 and 0.64. The Raman spectra were deconvoluted using a Gaussian-Lorentzian mixed profile as plotted in Figure 4(b). Composition dependence of the Raman shift is revealed in Figure 4(c). The $A_1$(1LO), and $E$(2TO) peaks between 200 cm$^{-1}$ and 450 cm$^{-1}$ , corresponding to the Ti-O and Sc-O bond vibrations [40]. A sudden change in Raman shifts of the two peaks around $x$ = 0.62 could be related to the phase transition from rhombohedral to tetragonal, which agrees with the XRD results.

Figure 5 presents SEM images of the BS-$x$PT-PCN ($x$ = 0.58-0.64) ceramics. The SEM photos exhibit that all the samples are dense with well-grown grain. The component of $x$ = 0.62 has the smallest grain size. Meanwhile the average grain size of the BS-$x$PT-PCN system is approximately 2.17~2.44 $\mu$m. The change of $x$ has no prominent impact on the grain size of the system. The dense and uniform morphology is fundamental for the ceramics to have excellent mechanical performances.

3.2 Dielectric behavior

Figure 6(a)-(d) show temperature as a function of dielectric performances ($\varepsilon_r$, tan$\delta$) of the BS-$x$PT-PCN ceramics. It can be obviously seen that the $T_c$ nearly linearly increases from 366 °C to 420 °C with the increase of $x$. For the MPB composition, the $T_C$ achieves 401 °C, which is a bit lower than that of pure BiScO$_3$-PbTiO$_3$ [29]. The loss tangent (tan$\delta$) of all specimens is lower than 0.08 from RT to 350 °C at 1 kHz, which illustrates that BS-$x$PT-PCN is potentially appropriate for harsh environment. The shape of dielectric permittivity peaks become suppressed and broader with the lower

PbTiO$_3$ contents, which belongs to diffuse phase transition behavior. The diffuse phase transformation is often ascribed to the balance breaking of component and structural disorder in the arrangement of non-isovalent Nb$^{5+}$ and Cd$^{2+}$ ions on the crystallographic equivalent sites.

For purpose of illustrate the dielectric relaxor behavior of the BS-$x$PT-PCN ($x$ = 0.58-0.64) ceramics, we used the quadratic Lorentz formula: [30]

$$\frac{\varepsilon_A}{\varepsilon} = 1 + \frac{(T-T_A)^2}{2\delta_A^2} \qquad (1)$$

where $T_A$ and $\varepsilon_A$ are the temperature of the dielectric peak and the speculative value of $\varepsilon_A$ at $T = T_A$, respectively. The coefficient $\delta_A$ represents the diffuseness of the dielectric peak. The greater is the relaxor dispersion, the larger is $\delta_A$. The temperature as a function of the dielectric permittivity at $T > T_m$ for BS-$x$PT-PCN ceramics was fitted by Eq. (1) as presented in Figure 6. As the fitting parameters (Table 3) indicate, it can be obviously seen that the parameter $\delta_A$ decreases as the PT content increased, implying the degree of diffusion phase transition decreases. which is related to a small amount of the Cd and Nb occuping the B site in the perovskite structure, leading to the B site ion distribution becoming disordered [31].

Frequency dispersion of the phase transition in ferroelectric materials can be described by the Vogel-Fulcher formula, which introduces the concept of "freezing temperature" and reflects the freezing process caused by the interaction between dipoles or polar nanoregions (PNRs). The relationship of the Vogel-Fulcher can be described as follows:

$$f = f_0 \exp(-E_a / k_B(T_m - T_f)) \qquad (2)$$

where $f_0$ is the frequency of tries to conquer the potential barrier $E_a$ ($f_0 \sim 10^{13}$, in our case); $k_B$ is the Boltzmann constant; $T_f$ is the static freezing temperature. Figure 7 presents the $T_m$ dependence of $f$. The obtained $E_a$, $T_f$ and $f_0$ are cataloged in Table 3. $T_f$ ascends from 590 to 678 K with PbTiO$_3$ increasing from 0.58 to 0.64, illustrating the presence of freezing behavior in the dipole dynamics upon RT [32]. The calculated $E_a$ is around 0.022~0.069 eV, $E_a$ gradually decreased with the increase of PT, indicating that the interaction between PNRs increased.

3.3 Piezoelectric and ferroelectric properties

The *P-E* hysteresis loops of the specimens are presented in Figure 8(a). All the polarized samples exhibit saturated *P-E* loops under driven fields. Figure 8(b) demonstrates the composition dependence of the coercive field ($E_c$) and the remnant polarization ($P_r$). With the increase of $x$, the $P_r$ initially increases, attains the maxima 40 μC/cm² around the MPB ($x$ = 0.62), then decreases to ~28 μC/cm² for $x$ = 0.64. This should be attributed to facilitated polarization by phase coexistence, where monoclinic phase equipped twenty four spontaneous polarization directions and six autonomous polarization directions for tetragonal phase, giving a sum thirty directions [33]. The highest ferroelectric property was obtained at the MPB content $x$ = 0.62 with a coercive field $E_c$ = 19.6 kV/cm and remnant polarization $P_r$ = 40 μC/cm². The $E_c$ first reduces slightly to 19.6 kV/cm, and then increases to 25.9 kV/cm. This is consistent with the XRD results since the domain switching getting harder with the increase fraction of the tetragonal phase. As exhibited in Table 2, Figure 2 and Figure 8, the change tendency of piezoelectric parameter $d_{33}$ is analogous to that of $P_r$ and $ε$, which can be revealed as follows: $d_{33}$ is proportional to $Q\varepsilon_0\varepsilon P_r$ [34], where $Q$, $\varepsilon_0$, $\varepsilon$ as well as $P_r$ represent the electrostriction coefficient, dielectric parameter of vacuum, dielectric permittivity as well as remnant polarization of samples, respectively.

3.4 Temperature as a function of depoling and phase structure of BS-$x$PT-PCN ceramics

Figure 9(a) and (b) show the thermal stability of $d_{33}$ and $k_p$ from RT to 400 °C of the BS-$x$PT-PCN system. The piezoelectric coefficient maintains decent stability, meanwhile, harsh deterioration of the piezoelectric properties does not happen until near the $T_c$. To further shed light on the connection between the deterioration of piezoelectric characterization and depoling temperature, an *in situ* XRD of the BS-0.62PT-PCN was accomplished from RT to 450 °C as plotted in Figure 10(a). Figure 10(b) presents the variation of the phase fraction of monoclinic, tetragonal and cubic dependence on temperature. In the temperature region RT to 250 °C, the BS-0.62PT-PCN mainly sustains its tetragonal phase, revealing high thermal stability of vertical MPB. Hence, the piezoelectric performances of BS-0.62PT-PCN present good temperature stability ascribed to the phase fraction maintains unchanged. It is consistent with the dielectric behavior and annealing test of $d_{33}$. Further increasing temperature, the appearance of cubic phase around 400 °C implies an evident phase transformation from ferroelectric to paraelectric happens. As a result, the $d_{33}$ and $k_p$ degenerates to zero around the $T_c$. Consequently, the reason of excellent piezoelectric performance of the BS-$x$PT-PCN ceramics around MPB is ascribed to the low symmetry monoclinic phase (where the switching of polarization is easy) while the decent high-temperature thermal stability is attributed to the good stability of tetragonal and monoclinic phases. The results validate that BS-$x$PT-PCN ceramics are a promising candidate for high temperature piezoelectric applications.

## 4. Conclusions

The BS-$x$PT-PCN ceramics (0.58 ≤ $x$ ≤ 0.64) were manufactured by the traditional solid state reaction method. The phase structure, microstructure as well as electrical performances of the BS-$x$PT-PCN ceramics were investigated. The increase of PT results in a structural phase transformation from rhombohedral phase to tetragonal phase. The vicinity of MPB ($x$ = 0.62) is demonstrated by a mixture of phases: monoclinic phase (space group, *Cm*) and tetragonal phase (space group, *P4mm*). The ceramic with $x$ = 0.62 indicates optimal dielectric, piezoelectric, and ferroelectric performances with $\varepsilon_{rmax}$ = 31565, $T_c$ = 401 °C, $d_{33}$ = 508 pC/N, $k_p$ = 56%, and $P_r$ = 40 μC/cm$^2$. With a higher $T$c than merchant PZT ceramics, BS-$x$PT-PCN ceramics exhibit excellent piezoelectric parameter of 508 pC/N, which is higher than $d_{33}$ (460 pC/N) of pure BS-PT ceramics. Thermal depoling on piezoelectric properties dependence on temperature indicates a decent temperature stability. It is considered that the BS-$x$PT-PCN ceramics are promising for application as high-temperature sensors and actuators.

## Acknowledgments

This work was financially supported by the Natural Science Foundation of China (NSFC Grant Nos. 11564010, 51402196, 51602159). D.W. acknowledges the support of NSFC (Grant Nos. 11574246 and U1537210) and the National Basic Research

Program of China, Grant No. 2015CB654903. L.L also thanks the support from the Natural Science Foundation of Guangxi (Grant No. GA139008, AA138162, CB380006, FA198015), the Research Grants Council of the Hong Kong Special Administrative Region, China (ProjectNo.PolyU152665/16E), the Scientific Research Foundation of Guangxi University (GrantXTZ160530).

**Table 1** Results of the Rietveld refinement of the sample with $x$ = 0.58, 0.60, 0.62 and 0.64 at RT.

| PbTiO$_3$ content | $x$ = 0.58 | $x$ = 0.60 | $x$ = 0.62 | | $x$ = 0.64 |
|---|---|---|---|---|---|
| Space group | R3m | R3m | P4mm | Cm | P4mm |
| Phase fraction | 1 | 1 | 0.732 | 0.268 | 1 |
| a | 4.03302(0) | 4.02739(3) | 3.99367(4) | 5.67819(2) | 3.99217(7) |
| b | 4.03302(0) | 4.02739(3) | 3.99367(4) | 5.66168(8) | 3.99217(7) |
| c | 4.03302(0) | 4.02739(3) | 4.06155(5) | 4.04328(1) | 4.07647(7) |
| β | 89.8176 | 89.8742 | 90.0000 | 89.8007 | 90.000 |
| Cell volume (Å$^3$) | 65.609 | 65.323 | 64.780 | 130.711 | 64.971 |
| Bi/Pb | | | | | |
|   x | 0.5 | 0.5 | 0 | 0 | 0 |
|   y | 0.5 | 0.5 | 0 | 0 | 0 |
|   z | 0.5 | 0.5 | 0 | 0 | 0 |
| Ti | | | | | |
|   x | 0.0466820 | 0.046700 | 0.5 | 0.480000 | 0.5 |
|   y | 0.0466820 | 0.046700 | 0.5 | 0 | 0.5 |
|   z | 0.0466820 | 0.046700 | 0.524000 | 0.656030 | 0.52368 |
| Sc | | | | | |
|   x | 0.046682 | 0.046700 | 0.5 | 0.480000 | 0.5 |
|   y | 0.046682 | 0.046700 | 0.5 | 0 | 0.5 |
|   z | 0.046682 | 0.046700 | 0.524000 | 0.656030 | 0.524000 |
| Nb | | | | | |
|   x | 0.046682 | 0.046700 | 0.5 | 0.480000 | 0.5 |
|   y | 0.046682 | 0.046700 | 0.5 | 0 | 0.5 |
|   z | 0.046682 | 0.046700 | 0.524000 | 0.656030 | 0.524000 |
| Cd | | | | | |
|   x | 0.046682 | 0.046700 | 0.5 | 0.480000 | 0.5 |
|   y | 0.046682 | 0.046700 | 0.5 | 0 | 0.5 |
|   z | 0.046682 | 0.046700 | 0.524000 | 0.656030 | 0.524000 |
| O1 | | | | | |
|   x | 0.552410 | 0.551200 | 0.5 | 0.457110 | 0.5 |
|   y | 0.065260 | 0.066500 | 0.5 | 0 | 0.5 |
|   z | 0.065260 | 0.066500 | 0.123000 | 0.031670 | 0.123000 |
| O2 | | | | | |
|   x | | | 0.5 | 0.225660 | 0.5 |
|   y | | | 0 | 0.252280 | 0 |

| | | | | | | |
|---|---|---|---|---|---|---|
| $z$ | | | 0.584000 | 0.543610 | | 0.584000 |
| $R_{wp}$ | 0.0640 | 0.0639 | 0.0635 | | | 0.0788 |
| $R_p$ | 0.0461 | 0.0452 | 0.0450 | | | 0.0515 |
| $x^2$ | 4.861 | 4.657 | 5.597 | | | 6.995 |

**Table 2** Room temperature dielectric and piezoelectric properties of BS-$x$PT-PCN ceramics.

| Composition | $\varepsilon_{rmax}$ | $\tan\delta$ | $T_c$ (°C) | $d_{33}$ (pC/N) | $K_p$ | $P_r$ (μC/cm$^2$) | $E_c$ (kV/cm) | |
|---|---|---|---|---|---|---|---|---|
| $x = 0.58$ | 16990 | 0.0313 | 366 | 256 | 0.41 | 28 | 21.7 | |
| $x = 0.60$ | 19103 | 0.0307 | 380 | 375 | 0.50 | 30 | 17.8 | |
| $x = 0.62$ | 31565 | 0.0289 | 401 | 508 | 0.56 | 40 | 19.6 | |
| $x = 0.64$ | 26562 | 0.0237 | 420 | 244 | 0.38 | 27 | 25.9 | |
| BS-0.42PT-0.42PMN | 11110 | 0.0660 | 102 | 509 | 0.46 | 28 | 12 | [35] |
| BS-0.64PT-0.03PMS | | 0.0075 | 365 | 300 | 0.43 | 22 | 17 | [36] |
| BS-0.58PT-0.06BST | | 0.05 | 384 | 440 | 0.46 | 29 | 20 | [37] |
| BS-0.61PT-0.03LN | | 0.0073 | 337 | 551 | 0.51 | 46.5 | 23 | [38] |
| BS-0.64PT | | 0.020 | 450 | 460 | 0.56 | 32 | 20 | [39] |

**Table 3** Fitting parameters for different compositions of BS-$x$PT-PCN solid solution obtained using Vogel-Fulcher law and quadratic Lorentz.

| Composition | $T_f$ (K) | $f_0$ (Hz) | $E_a$ (eV) | $T_A$ (K) | $\delta_A$ (K) | $\varepsilon_A$ |
|---|---|---|---|---|---|---|
| $x = 0.58$ | 590 | 1.50×10$^{13}$ | 0.069 | 611 | 88 | 15514 |
| $x = 0.60$ | 614 | 1.86×10$^{13}$ | 0.058 | 633 | 82 | 17599 |
| $x = 0.62$ | 630 | 1.79×10$^{13}$ | 0.054 | 647 | 64 | 28197 |
| $x = 0.64$ | 678 | $1.75\times10^{13}$ | 0.022 | 683 | 59 | 22904 |

**Figure Captions**

**Fig. 1** (a) XRD patterns (b) magnified XRD profiles of {200} of the BS-$x$PT-PCN ($0.58 \leq x \leq 0.64$) ceramics.

**Fig. 2** Piezoelectric coefficient $d_{33}$ and planar electromechanical coupling factor $k_p$ for BS-$x$PT-PCN at RT.

**Fig. 3** Rietveld refinement for the BS-$x$PT-PCN ($x$ = 0.58, 0.60, 0.62, 0.64) ceramics at RT.

**Fig. 4** (a) Room temperature Raman spectra, (b) Fitting spectra and (c) The change of peak positions based on the fitting results of Raman spectra for the BS-$x$PT-PCN system with different compositions.

**Fig. 5** SEM micrographs of the BS-$x$PT-PCN ceramics: (a) $x$ = 0.58; (b) $x$ = 0.60; (c) $x$ = 0.62; (d) $x$ = 0.64. Insert shows the grain size distribution of BS-$x$PT-PCN ceramics.

**Fig. 6** Temperature dependence of dielectric properties ($\varepsilon_r$, $tan\delta$) of the BS-$x$PT-PCN ($0.58 \leq x \leq 0.64$) ceramics. Plots of quadratic Lorentz fittings of the temperature dependence of the dielectric permittivity at 100 kHz.

**Fig. 7** The results of Vogel-Fulcher fitting for the BS-$x$PT-PCN ceramics with $x$ = 0.58, 0.60, 0.62, 0.64.

**Fig. 8** (a) Polarization hysteresis loops of the BS-$x$PT-PCN ceramics (b) The $P_r$ and $E_c$ of the ceramics with different $x$.

**Fig. 9** Effect of thermal depoling on the (a) piezoelectric parameter $d_{33}$ and (b) planar electromechanical coupling factor $k_p$ of the BS-$x$PT-PCN ceramics.

**Fig. 10** (a) Structural phase transition as a function of temperature for $x = 0.62$. A selected range of $2\theta$ shows the transformation of the {200} Bragg Reflection with the temperature. (b) tetragonal and monoclinic phase fractions as a function of temperature.

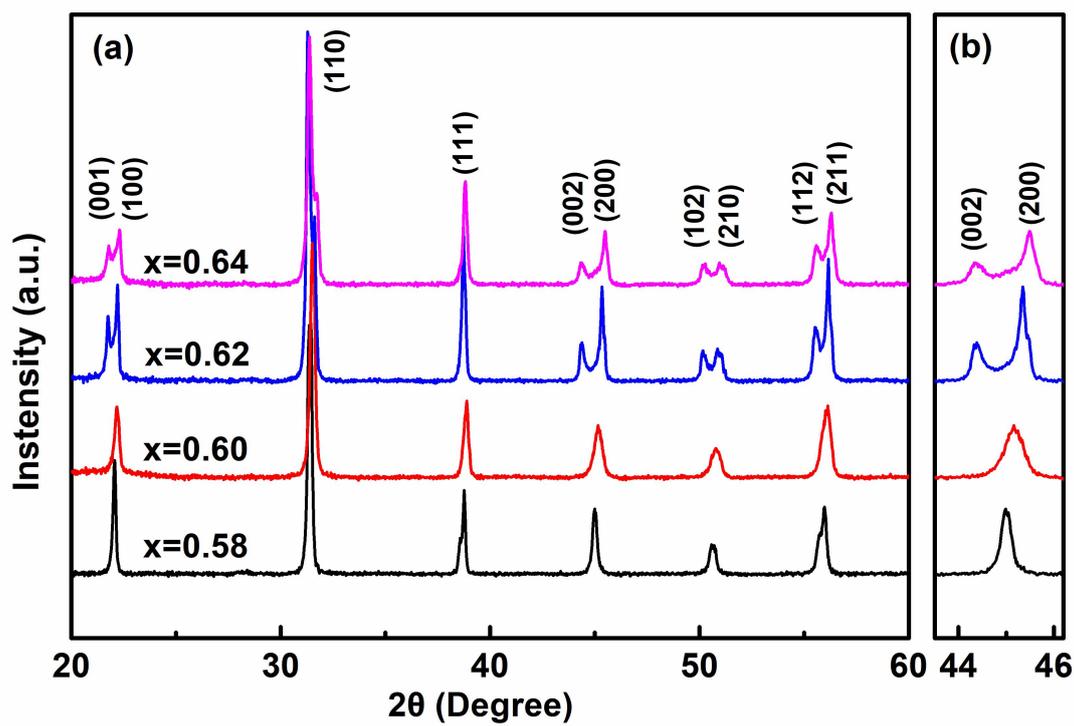

Fig. 1

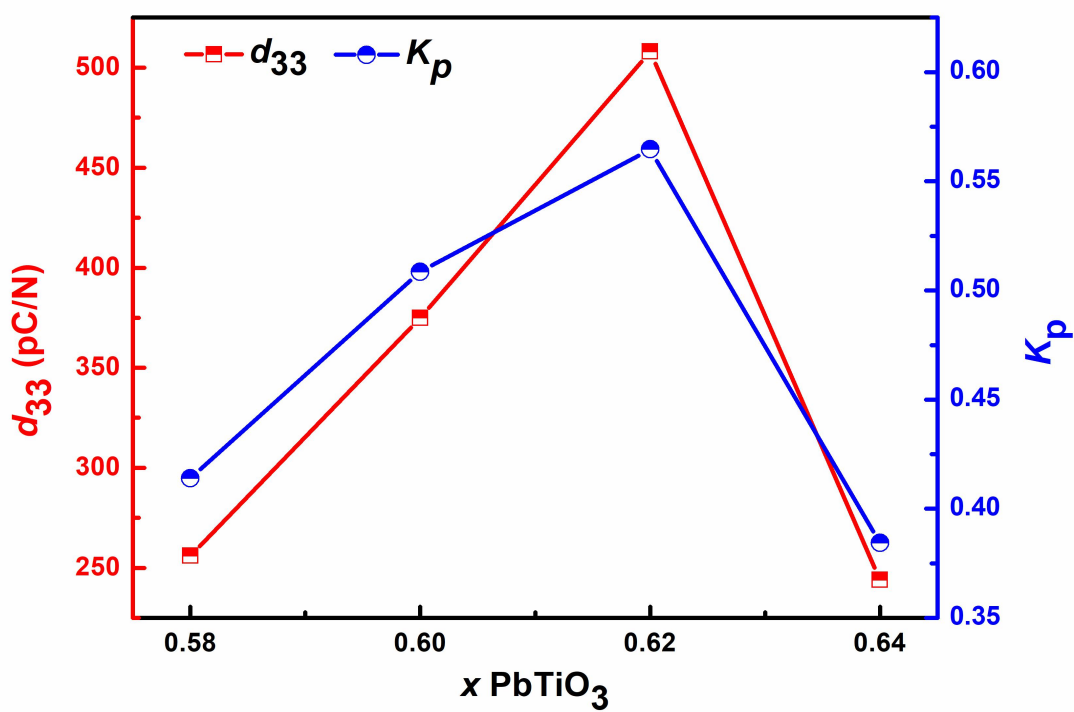

Fig. 2

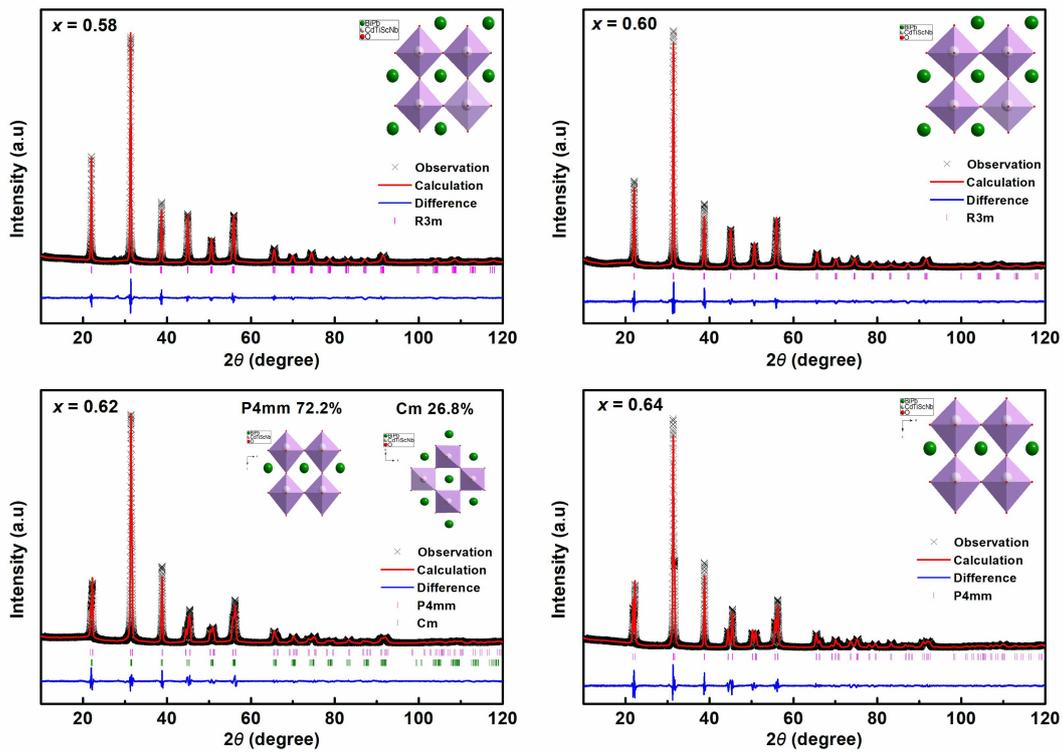

**Fig. 3**

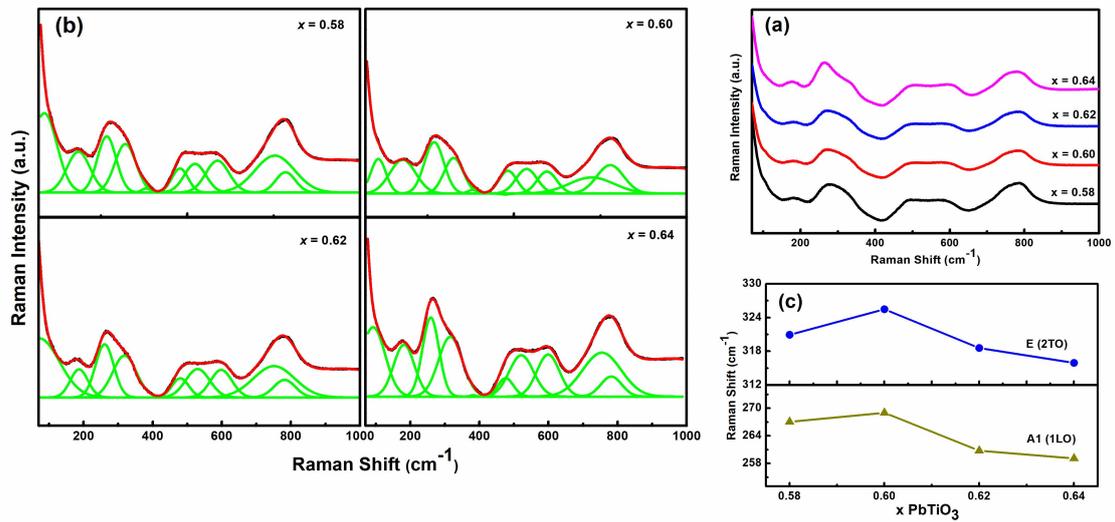

**Fig. 4**

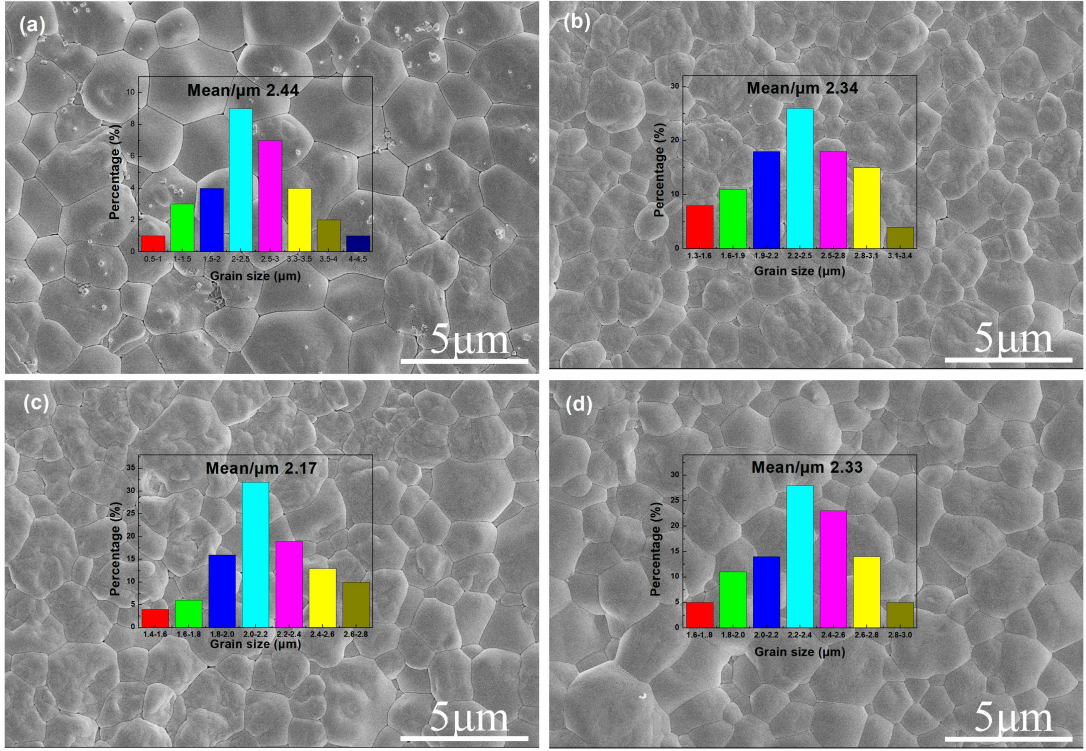

**Fig. 5**

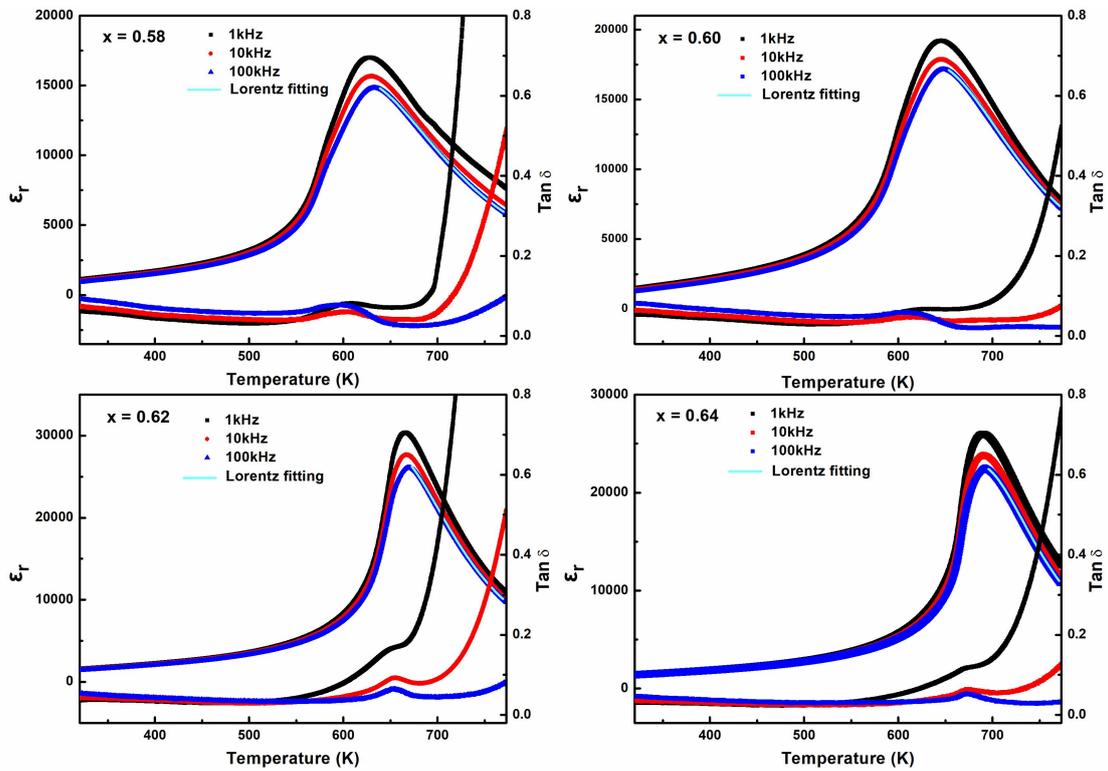

Fig. 6

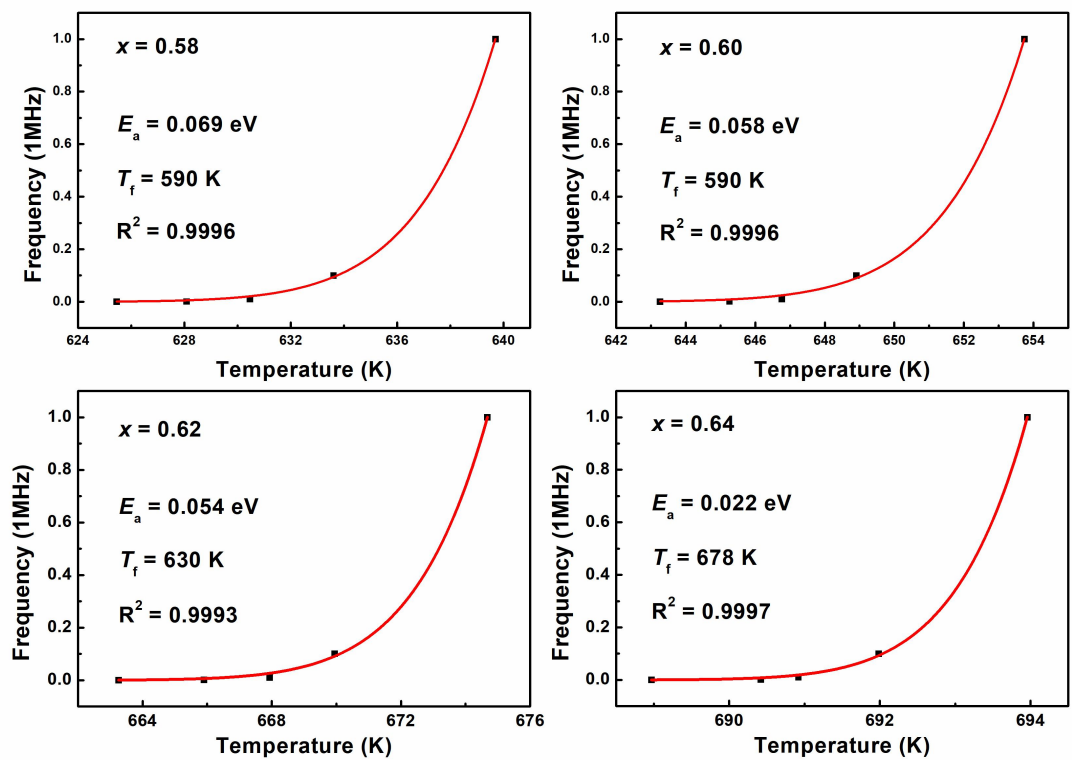

Fig. 7

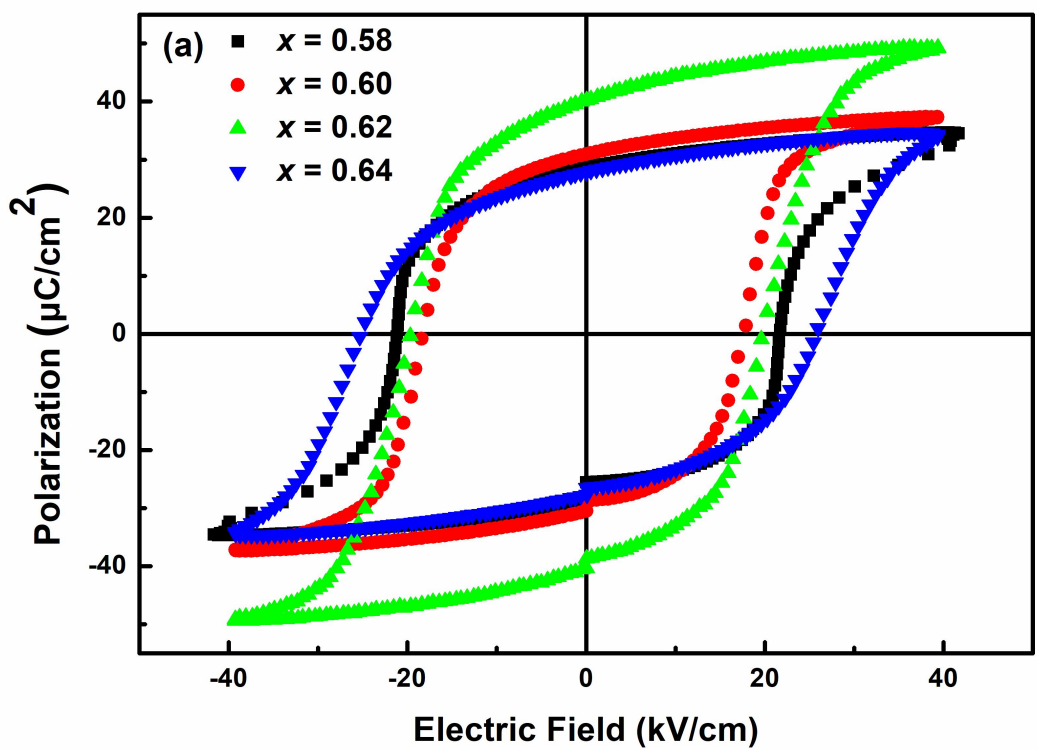

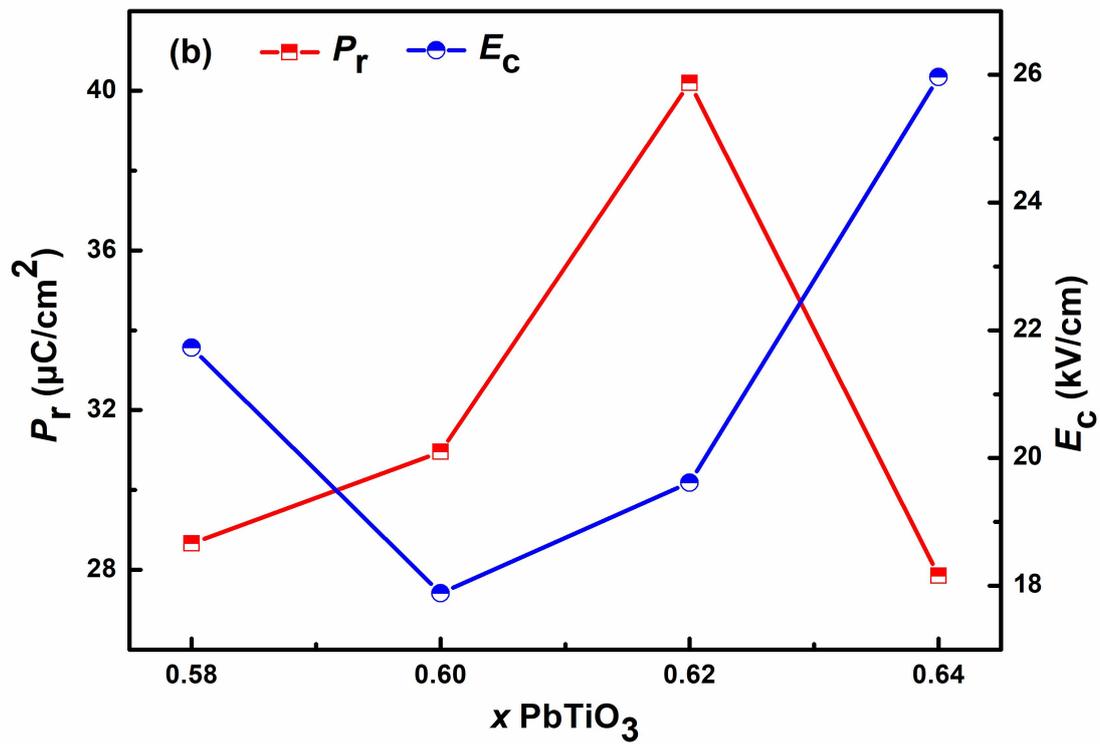

Fig. 8

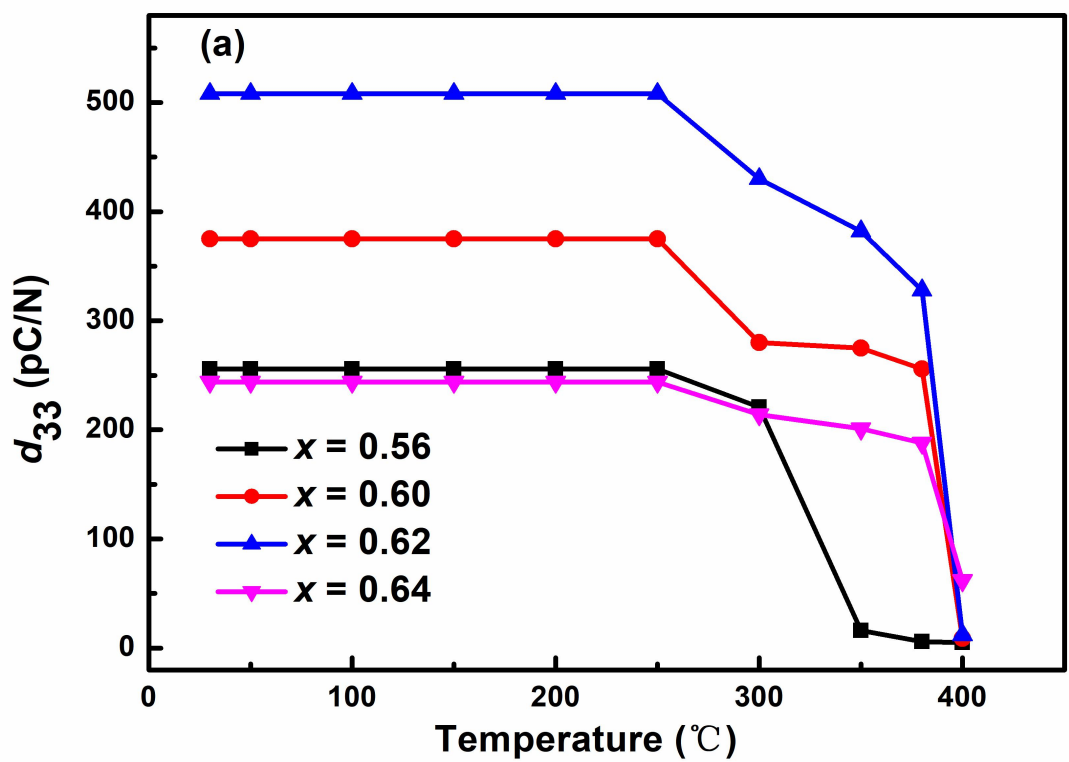

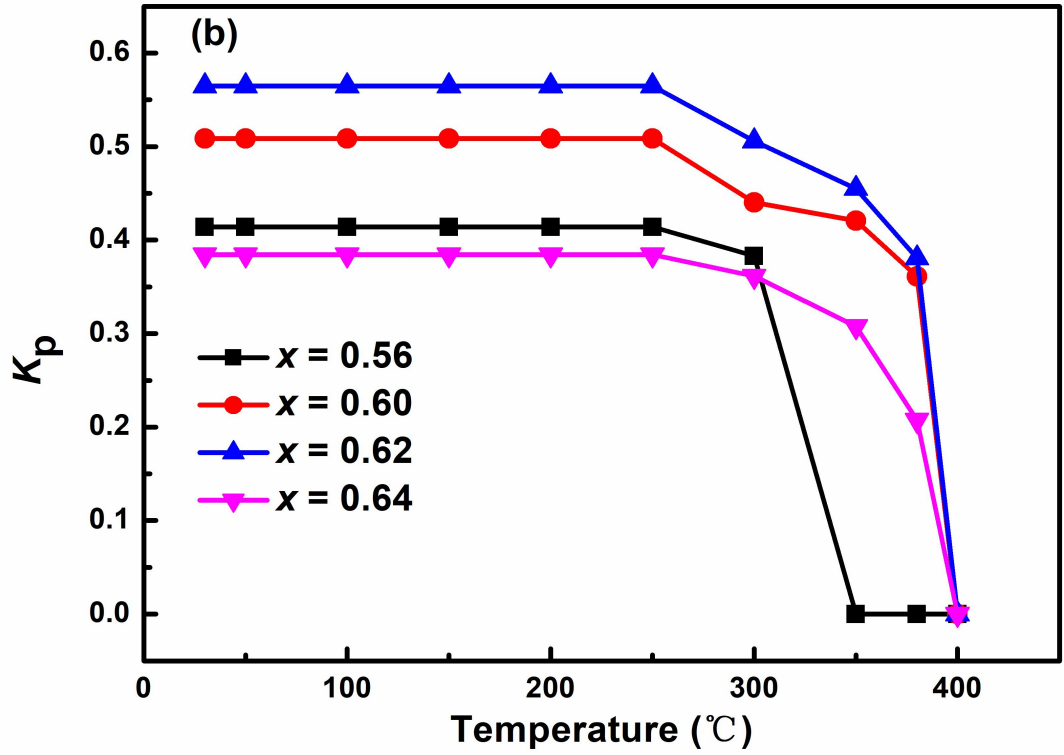

Fig. 9

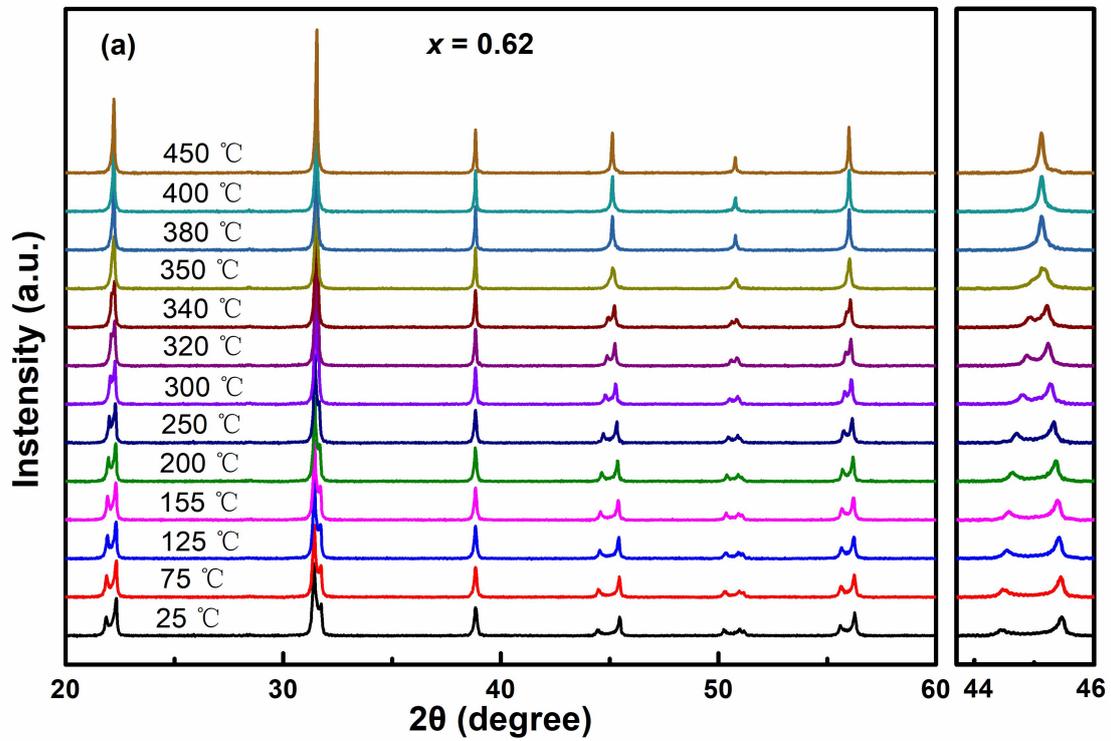

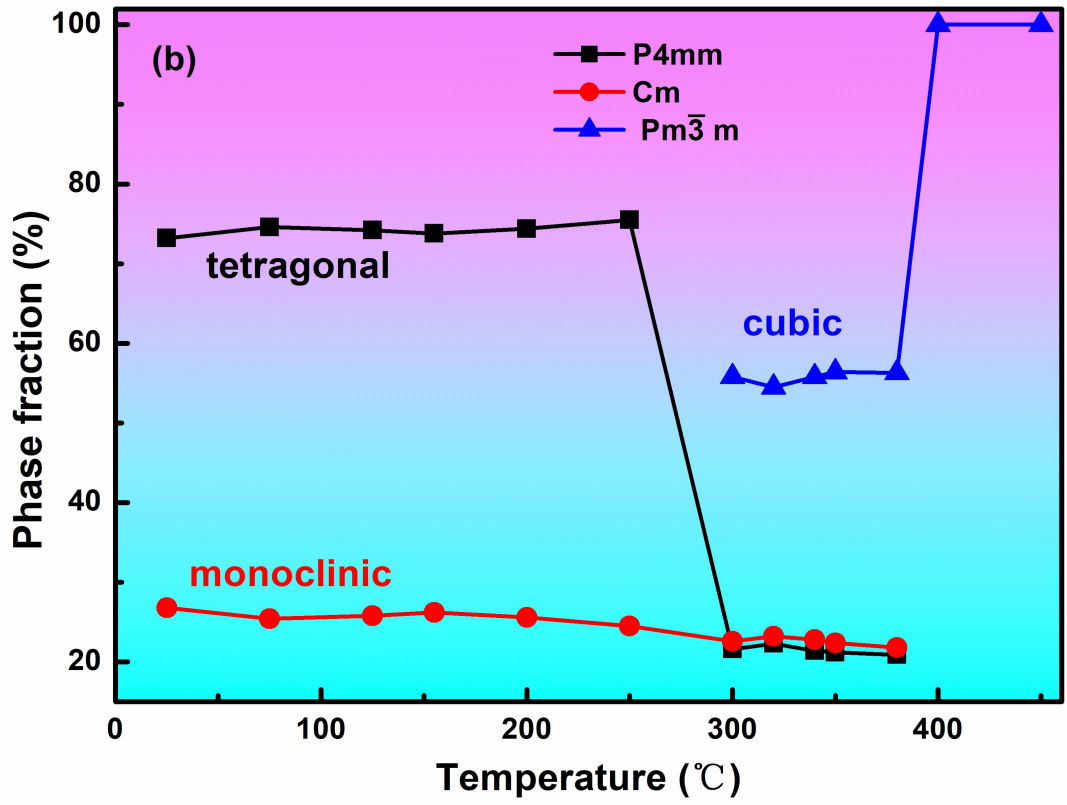

**Fig. 10**